\documentclass[10pt]{article}

\usepackage{amsmath}
\usepackage{amssymb}

\usepackage{graphicx}

\usepackage{cite}

\usepackage{color} 
\usepackage{CJKutf8}

\usepackage[labelfont=bf,labelsep=period,justification=raggedright]{caption}

\makeatletter
\renewcommand{\@biblabel}[1]{\quad#1.}
\makeatother

\date{}

\begin{document}

\begin{flushleft}
{\Large
\textbf{Robust identification of noncoding RNA from transcriptomes requires phylogenetically-informed sampling}
}
\\
Stinus Lindgreen$^{1,2,\mathbf{co}}$, 
Sinan U\u{g}ur Umu$^{2,3,\mathbf{co}}$, 
Alicia Sook-Wei Lai$^{2}$,
Hisham Eldai$^{2}$,
Wenting Liu$^{2}$,
Stephanie McGimpsey$^{2}$,
Nicole E. Wheeler$^{2}$,
Patrick J. Biggs$^{4,5}$,
Nick R. Thomson$^{6}$,
Lars Barquist$^{6,7}$,
Anthony M. Poole$^{2,3,5,\ast}$,
Paul P. Gardner$^{2,3,\ast}$
\\
{\bf 1} Department of Biology, University of Copenhagen, Copenhagen, Denmark.
\\
{\bf 2} School of Biological Sciences, University of Canterbury, Christchurch, New Zealand
\\
{\bf 3} Biomolecular Interaction Centre, University of Canterbury, Christchurch, New Zealand
\\
{\bf 4} Institute of Veterinary, Animal \& Biomedical Sciences, Massey University, Palmerston North, New Zealand
\\
{\bf 5} Allan Wilson Centre for Molecular Ecology \& Evolution, Massey University, Palmerston North, New Zealand
\\
{\bf 6} Pathogen Genetics, Wellcome Trust Sanger Institute, Hinxton, UK
\\
{\bf 7} Institute for Molecular Infection Biology, University of Wuerzburg, Wuerzburg, Germany
\\
{\bf co:} These authors contributed equally.
\\
$\ast$ Correspondence to A.M.P. or P.P.G.\\ 
E-mails: anthony.poole@canterbury.ac.nz; paul.gardner@canterbury.ac.nz
\end{flushleft}


\section*{Abstract}

Noncoding RNAs are integral to a wide range
of biological processes, including translation, gene regulation,
host-pathogen interactions and environmental sensing. While genomics
is now a mature field, our capacity to identify noncoding RNA elements
in bacterial and archaeal genomes is hampered by the difficulty of \emph{de
novo} identification. The emergence of new technologies for
characterizing transcriptome outputs, notably RNA-seq, are improving
noncoding RNA identification and expression quantification. However, a
major challenge is to robustly distinguish functional outputs from
transcriptional noise. To establish whether annotation of existing
transcriptome data has effectively captured all functional outputs, we
analysed over 400 publicly available RNA-seq datasets spanning 37 different
Archaea and Bacteria. Using comparative tools, we identify close to a
thousand highly-expressed candidate noncoding RNAs. However, our
analyses reveal that capacity to identify noncoding RNA outputs is
strongly dependent on phylogenetic sampling. Surprisingly, and in
stark contrast to protein-coding genes, the phylogenetic window for
effective use of comparative methods is perversely narrow: aggregating
public datasets only produced one phylogenetic cluster where these
tools could be used to robustly separate unannotated noncoding RNAs
from a null hypothesis of transcriptional noise. Our results show that
for the full potential of transcriptomics data to be realized, a
change in experimental design is paramount: effective transcriptomics
requires phylogeny-aware sampling.

\section*{Author Summary}

We have analysed over 400 public transcriptomes, generated using
RNA-seq, from almost 40 strains of Bacteria and Archaea.  We discovered
that the capacity to identify noncoding RNA outputs from this data is
strongly dependent on phylogenetic sampling.  Our results show that,
for the full potential of transcriptomics data as a discovery tool to
be realized, a change in experimental design is critical: effective
comparative transcriptomics requires phylogeny-aware sampling.

We also examined how comparative transcriptomics experiments can be
used to effectively identify RNA elements. We find that, for RNA
element discovery, a phylogeny-informed sampling approach is more
effective than analyses of individual species.  Phylogeny-informed
sampling reveals a narrow ‘Goldilocks Zone’ (where species are not too
similar and not too divergent) for RNA identification using clusters
of related species.

In stark contrast to protein-coding genes, not only is the
phylogenetic window for the effective use of comparative methods for
noncoding RNA identification perversely narrow, but few existing
datasets sit within this Goldilocks Zone: by aggregating public
datasets, we were only able to create one phylogenetic cluster where
comparative tools could be used to confidently separate unannotated
noncoding RNAs from transcriptional noise.

\section*{Introduction}

Genome sequencing has transformed microbiology, offering unprecedented
insight into the physiology, biochemistry, and genetics of Bacteria
and Archaea \cite{Wu:2009,Rinke:2013,Loman:2012,Chun:2014}. Equally,
careful examination of transcriptional outputs has revealed that
bacterial and archaeal transcriptomes are remarkably complex
\cite{Sorek:2011}. Roles for RNA include regulation, post-transcriptional
modification and genome defense processes
\cite{Storz:2011,Dennis:2001,Horvath:2010,Breaker:2012,Cech:2014}. However,
our view of the microbial RNA world still derives from a narrow
sampling of microbial diversity \cite{Pagani:2012}. Additional bias
comes from the fact that many microbes are not readily culturable
\cite{Stewart:2012}. The development of metagenomics and initiatives
such as the Genomic Encyclopedia of Bacteria and Archaea (GEBA)
project have sought to redress these biases, generating genomes
spanning undersampled regions of the bacterial and archaeal phylogeny
\cite{Wu:2009}, and sequencing uncultured or unculturable species
through metagenomics
\cite{Elkins:2008,Konneke:2005,Tyson:2004,Woyke:2006,Rinke:2013}.

A further source of bias in our genome-informed view of microbes
derives from a protein-centric approach to genome annotation. The
majority of genome sequences deposited in public databases carry
limited annotation of noncoding RNAs and cis-regulatory elements, yet
it is rapidly becoming clear that RNA is essential to our
understanding of molecular functioning in microbes \cite{mandin2013rna}. 

The paucity of annotations is understandable, as RNA gene annotation
is non-trivial \cite{Freyhult:2007,Nawrocki:2009}. However, the
increasing number of roles for RNAs uncovered through experimental and
bioinformatic studies make illuminating this ``dark matter'' all the more
urgent. Among the remarkable discoveries made are: riboswitch-mediated
regulation \cite{Barrick:2007,Breaker:2012}, transcriptional
termination by RNA elements
\cite{von_Hippel:1998,Gardner:2011,Santangelo:2011}, identification of
novel natural catalytic RNAs
\cite{Kruger:1982,Guerrier-Takada:1983,Winkler:2004,Roth:2014},
CRISPR-mediated acquired immunity \cite{Barrangou:2007,Brouns:2008},
temperature-dependent gene regulation \cite{Narberhaus:2006,Loh:2013},
and sno-like RNAs in Archaea
\cite{Omer:2000,Gaspin:2000,Gardner:2010}. The Rfam database
\cite{Gardner:2011,Burge:2013} provides a valuable platform for
collating and characterising these and other families of noncoding
RNA. However, a recent comparative analysis \cite{Hoeppner:2012}
revealed that fewer than 7\% of RNA families within Bacteria and less
than 19\% in Archaea show a broad phylogenetic distribution (that is,
presence in at least 50\% of sequenced phyla). Crucially, that
analysis revealed that underlying genome sequencing biases were a
major contributor to this pattern, and that the wider genomic sampling
provided by the GEBA dataset \cite{Wu:2009} did help improve
identification of broadly-conserved RNA families
\cite{Hoeppner:2012}. Tools such as RNA-seq \cite{Croucher:2010} and
transposon insertion sequencing
\cite{van_Opijnen:2013,Barquist:2013,Barquist:2013a} promise to
complement comparative genomics tools for RNA family discovery, and it
may be possible to use a mix of data types in the identification of
RNA elements. However, to date, no systematic analysis of available
data has been undertaken, suggesting ncRNAs may be hidden in the
deluge of published data.

We have therefore assessed the value of RNA-seq data for
identification of unannotated non-coding and cis-regulatory RNA
elements in bacterial and archaeal genomes. We show that numerous,
hitherto uncharacterised, expressed RNA families are lurking in
publicly available RNA-seq datasets. We find that poor sequence
conservation for RNA families limits the capacity to identify
evolutionarily conserved, expressed ncRNAs from existing genomic and
transcriptomic data. Our results suggest that maximising phylogenetic
distance, a sampling strategy effective for identification of novel
protein families \cite{Wu:2009,Rinke:2013}, is not the
most effective strategy for ncRNA identification. Instead, our results
show that, for RNA element identification, sequencing clusters of
related microbes will generate the greatest benefit.

\section*{Results}

\subsection*{Non-coding RNA elements dominate bacterial and archaeal transcriptional profiles}

To assess the relative contribution of noncoding RNAs and
protein-coding genes to transcriptional output, we collected all
publicly-available bacterial and archaeal RNA-seq datasets (available
as of August 2013), spanning 37 species/strains and 413 datasets. For
all datasets, we supplemented publicly available genome annotations
with screening for additional loci against the Pfam and Rfam databases
\cite{Punta:2012,Finn:2014,Gardner:2011,Burge:2013}, followed by
manual identification of expressed unannotated regions. This latter
annotation yielded 922 expressed RNAs of Unknown Function (RUFs)
\cite{McCutcheon:2003}.

\begin{figure}[!ht]
\begin{center}
\includegraphics[width=0.85\textwidth]{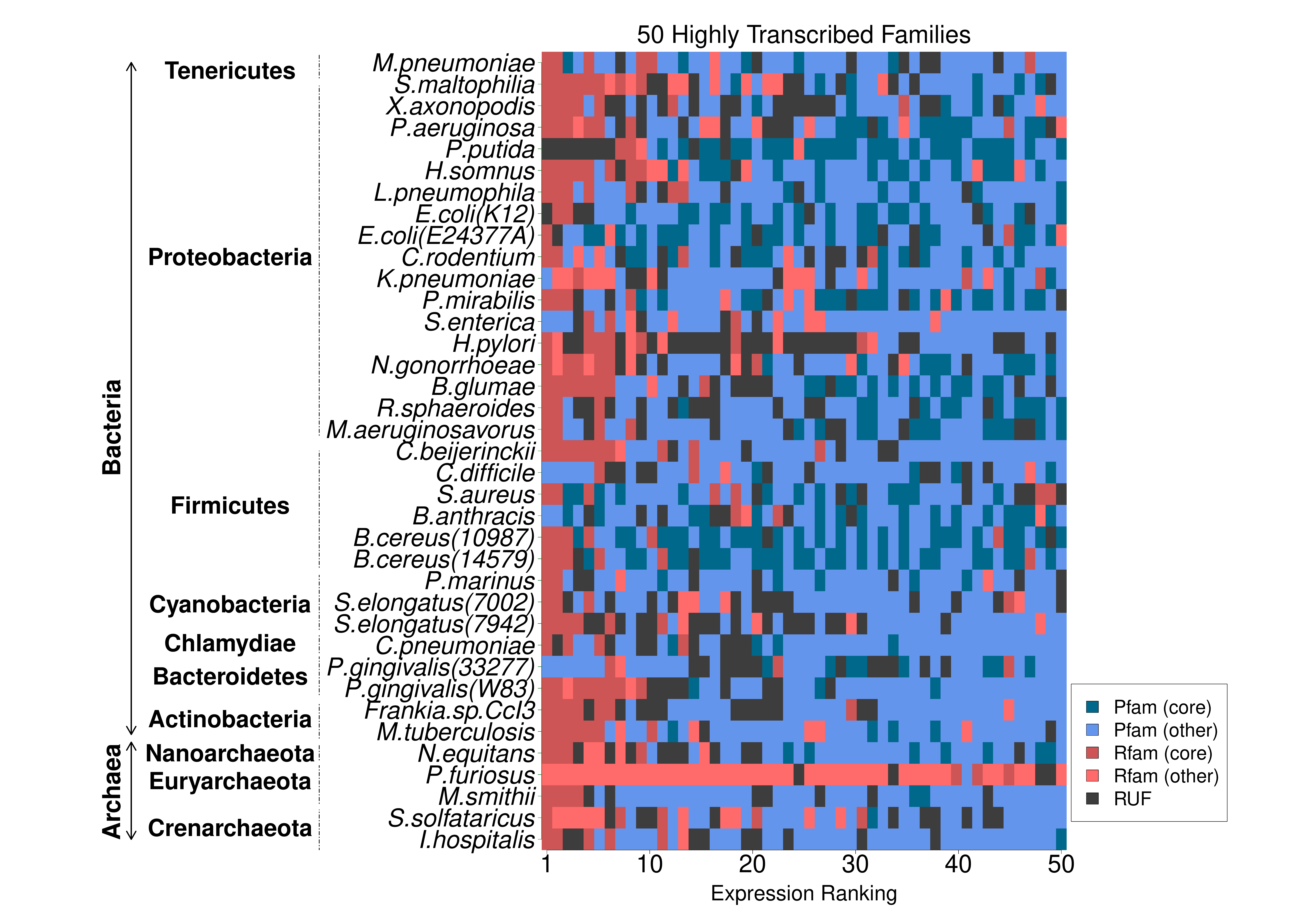}
\end{center}
\caption{ 
{\bf Identification of transcribed elements across
    publicly-available RNA-seq data.} Non-coding RNA elements show
  high expression across transcriptomes. Both annotated Rfam families
  (red - core Rfam families (see Methods) are dark red, all others are light red)
  and expressed RUFs (black) are among the highest expressed outputs
  in transcriptomes (blue - core Pfam families (see Methods) are dark blue, all
  others are light blue). For each strain we generated relative
  rankings of expression spanning protein coding genes, RNA genes and
  candidate RUFs. Accurately estimating expression levels from read depths is
  confounded by a number of factors (e.g. sample preparation, overall
  sequencing depths, rRNA depletion, etc.). For consistency, we
  have ranked genes for each strain and compared rankings instead
  of comparing the read depths directly between strains. For a given
  strain, the annotated genes were ranked based on the median read
  depth of the annotated region. RUFs were manually picked by masking
  out annotated genes and selecting regions showing evidence of
  expression by inspecting read depth across the genome. This yielded
  844 gene candidate sequences in Bacteria and 78 in Archaea. The plot
  contains the 50 most highly expressed elements for each
  strain/species.
}
\label{fig:1}
\end{figure}

We next examined the relative abundance of transcripts within each
RNA-seq dataset, yielding an expression rank for individual
transcripts. This analysis reveals that most transcriptomes are
dominated by highly expressed non-coding RNA outputs
(Figure~\ref{fig:1}) (P-value $<< 0.0001$, Chi-square test of observed
vs. expected ratios and Fisher’s Exact test on the counts). In
addition to well-characterised RNAs (rRNA, tRNA, tmRNA, RNase P RNA,
SRP RNA, 6S and sno-like sRNAs), and known cis-regulatory elements
(riboswitches, leaders and thermosensors - Table S1), the top 50
most abundant transcriptional outputs (Figure~\ref{fig:1}) across the
32 Bacteria and 5 Archaea in our dataset included a total of 308 RUFs.

\subsection*{Comparative analyses reveal that highly expressed transcripts are often poorly conserved}

\begin{figure}[!ht]
\begin{center}
\includegraphics[width=0.75\textwidth]{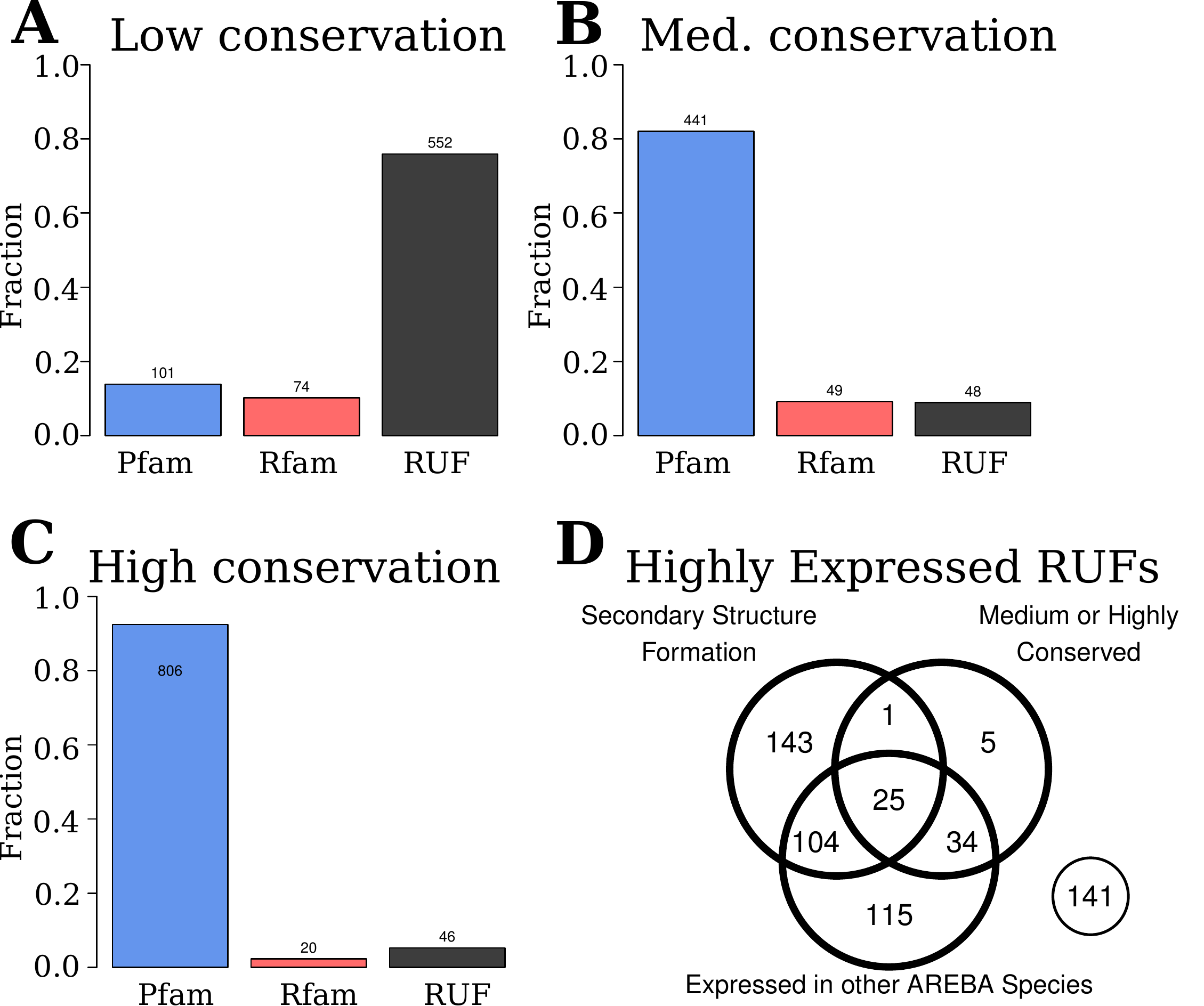}
\end{center}
\caption{ {\bf Many ncRNAs and RUFs are highly expressed but show
    limited conservation across represented strains/species.}
  {\bf A-C:} We have defined the ``family conservation'' for
    Pfam, Rfam and RUFs based upon the maximum phylogenetic distance
    (using structural SSU rRNA alignments) between any two strains
    hosting the family. We have divided the highly expressed
    transcripts (ranks 1-204) into Low, Medium and High conservation
    groups based on the lower-quartile, inter-quartile range and the
    upper-quartile of the family conservation measure (see Methods for
    further details). Both the known Rfam families and the RUFs
  identified in this analysis are often highly expressed
  transcripts. In contrast to protein-coding transcripts (blue), where
  highly-expressed transcripts are well-conserved, the opposite is
  true of many non-coding RNA elements (Rfam, red; RUFs,
  black). Notably, the greatest proportion of highly expressed
  Rfam-annotated RNA elements show a narrow evolutionary
  distribution. This is also reflected in the RUFs identified in this
  study. {\bf D:} Venn diagram of the 568 highly expressed RUFs. Each
  RUF was analysed to look for evidence of secondary structure
  formation, level of conservation, and evidence of expression in at
  least one other RNA-seq dataset. All RUFs showing expression in
  other strains/species are conserved in at least two strains/species,
  so the figure also shows that 219 highly expressed RUFs are
  conserved across a limited phylogenetic distance only.}
\label{fig:2}
\end{figure}

To assess whether highly expressed RUFs possess features commonly
associated with function, we employed three criteria: 1) evolutionary
conservation, 2) conservation of secondary structure, 3) evidence of
expression in more than one RNA-seq dataset. For this analysis, we
compared and ranked transcriptional outputs across species/strains (see
Methods for details). Based on the relative rank across RNA-seq
datasets and the maximum phylogenetic distance observed across all
genomes, each transcript was classified as high, medium or low
expression, and high, medium or low conservation. This yielded a set
of highly expressed transcripts consisting of 162 Rfam families, 568
RUFs and 1429 Pfam families. As expected
\cite{Rocha:2003,Pal:2001,Drummond:2005}, conserved, highly expressed
outputs are dominated by protein-coding transcripts
(Figure~\ref{fig:2}B\&C). In contrast, transcripts that are highly
expressed but poorly conserved are primarily RUFs
(Figure~\ref{fig:2}A). Of the 568 RUFs identified, only 25 are
supported by all three conservative criteria (conservation, secondary
structure and expression) (Figure~\ref{fig:2}D), a further 138 RUFs
are supported by two criteria (Figure~\ref{fig:2}D). Consequently, on
these criteria, the vast majority of RUFs appear indistinguishable
from transcriptional noise. However, as these RUFs are among the
mostly highly expressed transcripts in public RNA-seq data, we next
considered whether our criteria were sufficiently discriminatory to
identify functional RNAs. It is well established that not all
functional RNAs exhibit conserved secondary structure – antisense base
pairing with a target is common, and does not require intramolecular
folding \cite{Gottesman:2011}. This indicates that criterion 2 will
apply to some, but not all functional RNA elements. Criteria 1 and 3
both derive from comparative analysis: criterion 1 requires an
expressed RUF to be conserved in some other genome, while criterion 2
requires an expressed RUF to be expressed in another of the datasets
in our study. We therefore sought to examine how effective our
comparative analyses are given that the available data represent a
small sample (transcriptomes from 37 strains) and given that biases in
genome sampling across bacterial and archaeal diversity impact
comparative analysis of RNAs \cite{Hoeppner:2012}.

\subsection*{Comparative analysis reveals a 'Goldilocks Zone' for ncRNA identification}

Effective comparative analysis requires appropriate phylogenetic
distances between species under investigation \cite{Eddy:2005}.  For
discovery of protein-coding gene families, maximising phylogenetic
diversity across the tree of life has proven very effective
\cite{Wu:2009,Rinke:2013}. For non-coding RNA, underlying biases in
genome sampling do affect the assessment of ncRNA conservation, and
adding phylogenetic diversity improves the identification of broadly
conserved ncRNA families \cite{Hoeppner:2012}. However, few ncRNAs
appear conserved across broad evolutionary distances
\cite{Hoeppner:2012}.
We have therefore considered how species selection impacts comparative
analysis as a tool for the identification of conserved ncRNAs.

To assess the effect of strain selection on our capacity to identify
RNA families using comparative analysis, we first generated F84
phylogenetic distances between 2562 bacterial strains and 154 archaeal
strains using SSU rRNA sequences from each strain (see Methods for
details). Next, for each Rfam RNA family and Pfam protein family, we
identified the maximum phylogenetic distance between any two species/strains
that encode a given family. We then calculated the fraction of
conserved RNA and protein families for a given phylogenetic distance.

\begin{figure}[!ht]
\begin{center}
\includegraphics[width=0.75\textwidth]{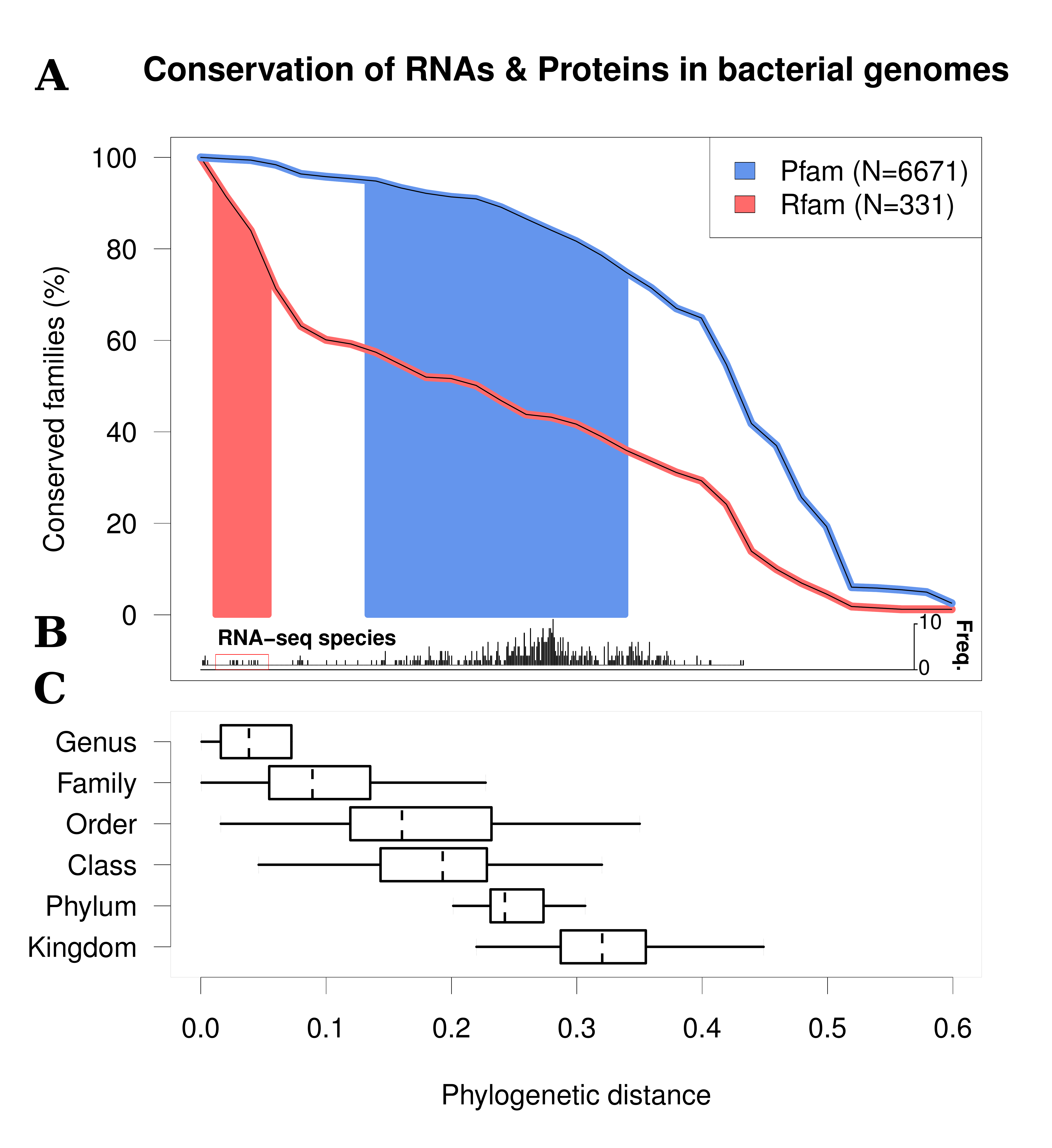}
\end{center}
\caption{ {\bf Conservation of protein and RNA families.}
  All of the available full length Bacterial and Archaeal
    genomes were annotated using Rfam and Pfam models. For each
    Pfam/Rfam family, RNA-seq species or taxonomic group the
    ``phylogenetic distance'' is calculated using the maximum SSU rRNA
    F84 distance (see Methods for details). {\bf A.} For the Pfam and the
    Rfam families we compare the levels of conservation as a function
    of phylogenetic distance using annotations of 2,562 bacterial
    genomes. E.g. $\approx 60\%$ of RNA families are conserved between
    species from the same family, whereas $>90\%$ of protein families
    are conserved within the same taxonomic range.  {\bf B.} The
    barplot shows the distribution of all pairwise distances between
    the RNA-seq datasets. Eleven pairs (boxed) are in the Goldilocks
    Zone (See Figure~\ref{fig:4} for further analysis). {\bf C.} The
    ranges of phylogenetic distances for comparing species from
    different taxonomic groups.}
\label{fig:3}
\end{figure}

This reveals a dramatic difference in evolutionary conservation of
Rfam and Pfam families (Figure~\ref{fig:3}). While 80\% of protein families are
still conserved at the broad evolutionary distances that separate
Bacteria and Archaea, the phylogenetic distance at which 80\% of RNA
families are conserved lies somewhere between the taxonomic levels of
genus and family (Figure~\ref{fig:3}). The explanation for this rapid decay of RNA
family conservation across long evolutionary time-scales is likely to
be a combination of the limited abilities of existing bioinformatic
tools to correctly align RNA sequences \cite{Gardner:2005} and 
rapid turnover of non-coding RNAs during evolution
\cite{Hoeppner:2012}.

These results in turn indicate that appropriate evolutionary distances
for optimal comparative analysis differ greatly for protein- and
RNA-coding genes. Figure~\ref{fig:3} confirms the utility of the GEBA sampling
strategy \cite{Wu:2009,Rinke:2013} for protein-coding gene
identification, since maximising phylogenetic diversity permits effective
identification of conserved protein-coding genes. In contrast, at the
largest phylogenetic distances, less than 40\% of the RNA families are
amenable to comparative analysis. These results define a ‘Goldilocks
Zone’ (an evolutionary distance neither too close nor too distant) for
ncRNA analysis through comparative analysis.

\begin{figure}[!ht]
\begin{center}
\includegraphics[width=0.99\textwidth]{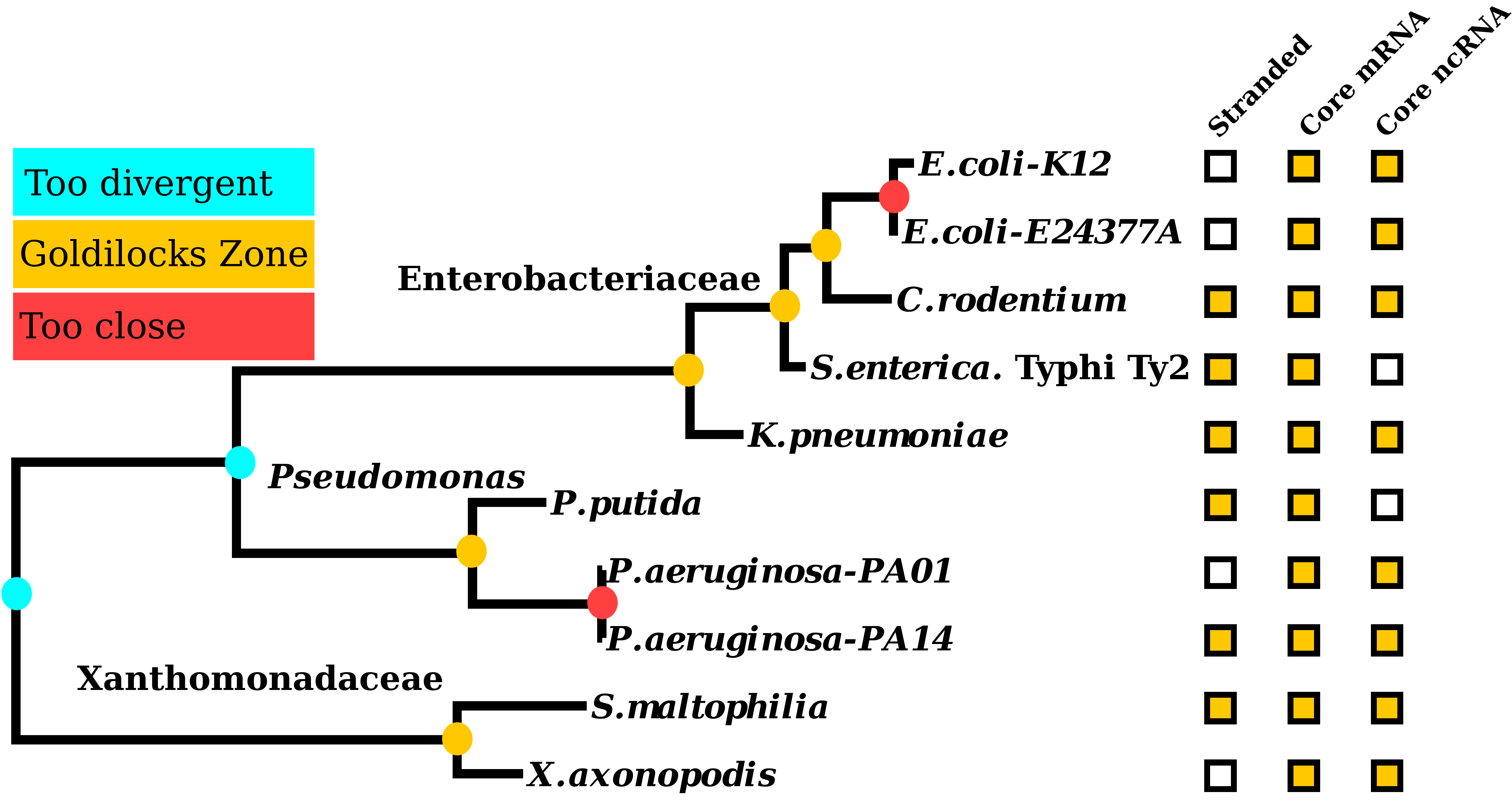}
\end{center}
\caption{ {\bf Public RNA-seq datasets that lie in the
      Goldilocks Zone.} Ten strains with corresponding, publicly
    available RNA-seq data and phylogenetic distances in the
    Goldilocks Zone (Figure 3) have been identified. The maximum
    likelihood tree from a SSU rRNA alignment shows the relationships
    between these taxa. They fall into three clades, containing
    members of the families: Enterobacteriaceae and Xanthomonadaceae,
    and the genus: \emph{Pseudomonas}. The nodes connecting taxa
    within the Goldilocks Zone are coloured gold, taxa that are too
    close are coloured red and those that are too divergent are
    coloured cyan. Each strain is annotated with gold boxes where
    there was stranded information, or if the majority of core mRNAs
    and ncRNAs (see Methods) were expressed (see Table S3 for the raw
    data). }
\label{fig:4}
\end{figure}

In order to assess the potential for existing RNA-seq data to be used
for ncRNA analysis, we mapped the pairwise distances between strains
covered by the RNA-seq datasets in this study. Of the 506 possible
pairs (excluding Bacteria vs Archaea), only 11 are in the Goldilocks
Zone for RNA (phylogenetic distance between 0.0118 and 0.0542)
covering 9 species/strains. While five pairs of datasets are ‘too hot’
(i.e. too close phylogenetically), the remaining 490 comparisons are
‘too cold’ for effective comparative RNA analysis
(Figure~\ref{fig:3}). The datasets in the Goldilocks Zone span three
distinct clades covering five Enterobacteria, three Pseudomanada, and
two Xanthomonada (Figure~\ref{fig:4}).

\begin{figure}[!ht]
\begin{center}
\includegraphics[width=0.99\textwidth]{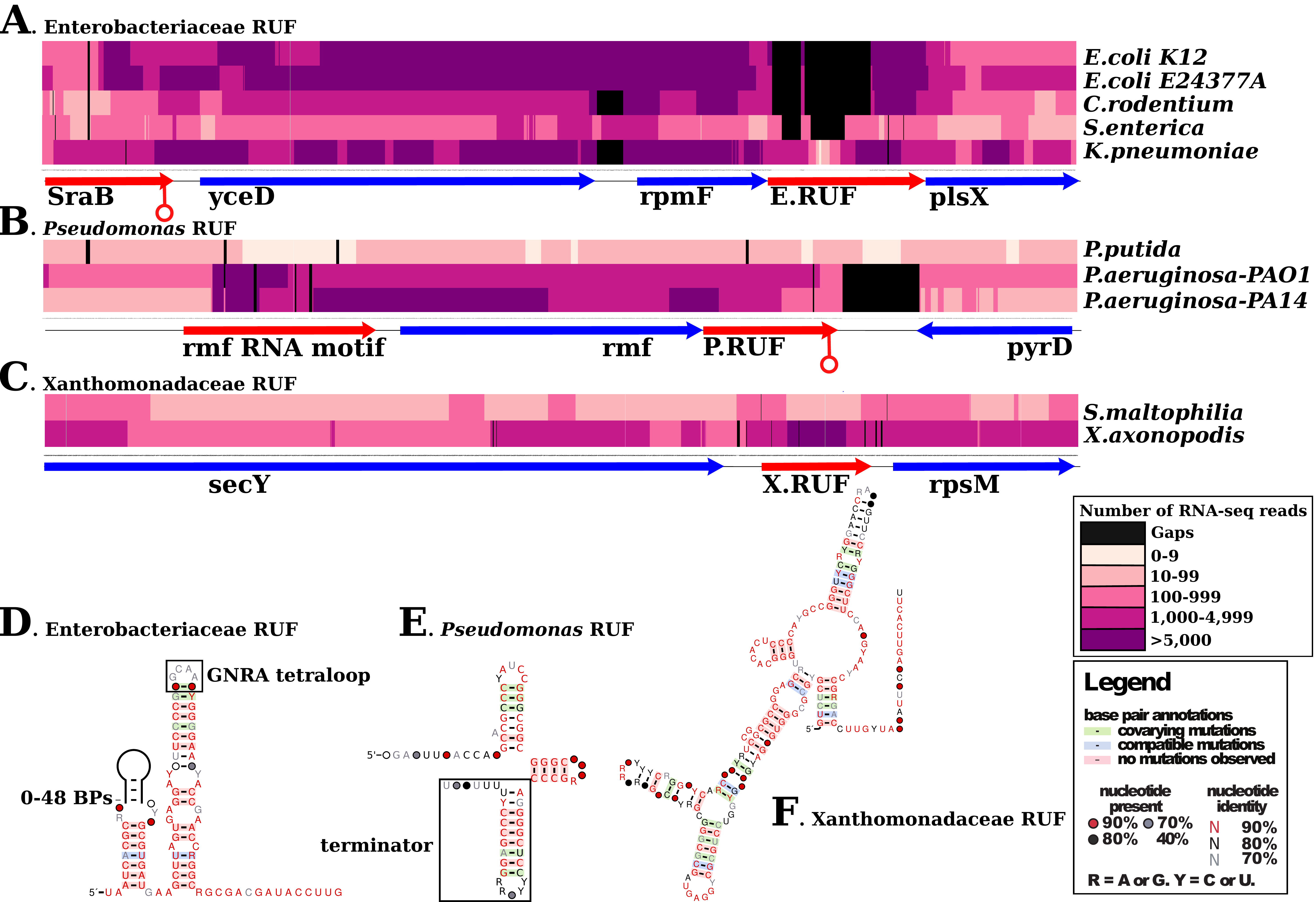}
\end{center}
\caption{ {\bf Comparative analysis of RNA-seq datasets in
      the Goldilocks Zone is a powerful approach for identifying
      RUFs.} In this figure we illustrate data corresponding to 3
    examplar RUFs that show high covariation, conserved predicted
    secondary structures and are derived from one of the Goldilocks
    Zone clades shown in Figure 4.  {\bf (A-C)} The expression levels
    inferred from RNA-seq in the region encompassing each RUF. The
    regions contain a mix of ncRNAs (red arrows) and protein coding
    genes (blue arrows) and a RUF (red arrow). For each nucleotide,
    the total number of reads that map to that nucleotide was
    computed, and are presented as a heatmap; darker colours indicate
    high relative expression, lighter colours indicate low expression
    and black indicates a gap in the genomic alignment of the
    sequences for the locii. {\bf (D-F)} R2R \cite{Weinberg:2011}
    representations of the predicted consensus secondary structures
    for exemplar RNAs of Unknown Function (RUFs) selected from the
    Enterobacteriaceae, \emph{Pseudomonas} and Xanthomonadaceae
    data. Covariation is highlighted in green, structure-neutral
    variation is highlighted in blue, highly conserved regions are
    highlighted in pink. The Enterobacteriaceae RUF contains a
    conserved tetraloop of the GNRA or UNCG type, and there have been
    two independent insertions of hairpins in \emph{S. enterica} and
    \emph{K. pneumoniae} within the first hairpin. The
    \emph{Pseudomonas} RUF hosts a 3$^\prime$ rho independent
    transcription terminator.} 
\label{fig:5}
\end{figure}

We next calculated the percentage of conserved RUFs for all
Enterobacterial strain pairs. On average, 83\% of RUFs are conserved
across the Goldilocks Zone. The two \emph{E. coli} strains are
extremely similar, and share 99\% of their RUFs, suggesting that these
strains are too similar for us to robustly separate expression of
\emph{bona fide} RNAs from noise. While these outputs could be genuine
RNAs, these strains are in the ‘too hot’ region, meaning if everything
is conserved, comparative power is lost. In contrast, only 12\% of
RUFs are conserved between strains/species pairs in the ‘too cold’
region (spanning clades; Figure~\ref{fig:4}) and of the 197 RUFs found
through comparative analysis of transcriptomes within the Goldilocks
Zone, only 19 show evidence of expression in another transcriptome
outside of this zone. This suggests that the low number of RUFs from
Figure~\ref{fig:2}D showing both conservation and expression is
primarily a consequence of limited sampling. That said, mining RNA-seq
data within the Goldilocks Zone permits a higher confidence in the
identification of novel ncRNAs. Three examples of this are
illustrated in Figure~\ref{fig:5}. These RUFs exhibit
sequence and secondary structure conservation and are expressed at high
levels across multiple Goldilocks Zone transcriptomes.

In summary, the Goldilocks Zone for RNA is surprisingly narrow, and
suggests that optimal strain selection for RNA comparative analyses
should comprise strains of the same species, members of the same
genus, and closely related taxonomic families (Figure~\ref{fig:3}). Thus, the
Goldilocks Zone for RNA is not encompassed by the sampling regimes
currently being employed for protein family discovery.

\section*{Discussion}

Our analyses of over 400 publicly-available bacterial and archaeal
RNA-seq datasets reveal that there is evidence for large numbers of RNAs of
unknown function in public data. We find evidence for close
to 1000 unannotated noncoding transcriptional outputs, but, given that
RNA-seq experiments provide a snapshot of gene expression under
specific experimental conditions, this number is likely to be far
lower than the complete set of transcriptional outputs. Thus, the
dataset we assembled for this project, which includes data generated
by a number of labs and derives from various species and strains grown
under a range of experimental conditions, is expected to represent a
broad, though partial, census of total expression outputs across the
species represented. Equally striking is the fact that, for the 922
RUFs identified in our study, over half (568) are among the most
abundant transcripts. These results suggest that ncRNA may play an
even greater role in the molecular workings of Bacteria and Archaea
than hitherto realised.

This use of transcriptome data clearly improves our capacity to
identify noncoding outputs: applying three criteria (sequence
conservation, conservation of secondary structure, and expression in
multiple strains/species) we have identified 163 high-confidence expressed
RUFs from public data (Figure~\ref{fig:2}). An additional 405 RUFs are highly
expressed across the transcriptomes we have examined, yet these do not
show clear signs of sequence or structural conservation in other
sequenced genomes. Given their high expression level, these seem
unlikely to be transcriptional noise. Some may represent technical
artefacts, but many could be \emph{bona fide} lineage-specific ncRNAs with
potentially novel functions.

Our results indicate that the greatest gain in analytical power for
ncRNA discovery will come from phylogenetically-informed experimental
design. Indeed, we find that this is critical to successful element
identification, since the ‘Goldilocks Zone’ for optimal comparative
analysis of RNA elements is surprisingly narrow. Hence, existing
efforts to maximise phylogenetic coverage of genome space
\cite{Wu:2009,Rinke:2013} need to be complemented with fine-scale
sampling of the tips (Figure~\ref{fig:4}). Indeed, analysing the few
transcriptomes that span the Goldilocks Zone reveals a remarkable
enrichment of transcripts showing evidence of structure, conservation
and expression in other strains/species. Furthermore, it is
  worth noting that the RNA family conservation decays as the
  phylogenetic distance increases (shown in Figure~\ref{fig:3}).
  There is a possibility that the Rfam families used for this are
  biased. However, if a bias exists, it is towards families with
  higher conservation (as the families are constructed from published
  ncRNAs that are often discovered based upon sequence conservation
  \cite{Gardner:2011,Burge:2013}). Thus, we might actually be
  overestimating RNA element conservation, making phylogenetically
  informed sampling even more important.

\begin{CJK}{UTF8}{min}
Given that isolation, cultivation and study of individual bacterial
and archaeal strains can be extremely challenging \cite{Stewart:2012}
successful phylogeny-informed comparative RNA-seq will be a demanding
endeavour, requiring complex sets of expertise spanning advanced
culturing and isolation techniques, functional genomics capability and
RNA bioinformatics. This places such a project beyond the reach of
most individual labs. We therefore propose that comprehensive
resolution of the comparative RNA-seq problem can best be resolved via
a community-driven initiative: in recognition of the success of the
GEBA project, we have dubbed this An RNA Encyclopedia of Bacteria and
Archaea (AREBA). The appropriateness of this acronym will be
especially clear to Japanophones, as, in Japanese, the phrase ‘areba’
(あれば) translates to ‘if there’.
\end{CJK}

\clearpage

\section*{Materials and Methods}

\subsection*{Preprocessing and mapping}

All available bacterial and archaeal genomes were downloaded from the
European Nucleotide Archive (ENA) (2,562 and 154 genomes,
respectively) \cite{Cochrane:2013}. RNA-seq datasets published as of
August 2013 were collected, spanning 37 species/strains, 44 experiments
and 413 lanes of sequencing data (Table S2). Most of these datasets
were generated on the Illumina platform \cite{Shendure:2005}, with a
few lanes from the SOLiD platform \cite{Cloonan:2008} and the 454
platform \cite{Margulies:2005}. Where possible, FastQ files were
downloaded, scanned for residual adapter sequences using
AdapterRemoval (v1.5.4) \cite{Lindgreen:2012}, and mapped to the reference genome
using Bowtie2 (v2.1.0) \cite{Langmead:2012} for Illumina and 454 data
and BFAST (v0.7.0a) \cite{Homer:2009} for SOLiD data.

\subsection*{Producing consistent genome annotations}

All genomes were re-annotated for both RNA genes and protein coding
genes. Non-coding RNA genes were annotated using cmsearch (v.1rc4)
\cite{Nawrocki:2013} to identify homologs of RNA families from the
Rfam database (v11.0) using the default ``gathering threshold''
(cmsearch --cut\_ga) \cite{Gardner:2011,Burge:2013}. Protein coding
genes were annotated using three approaches: First, annotations were
parsed from the ENA files. Secondly, Glimmer (v3.02) was run on all
genomes to predict open reading frames (with parameters ``-o7 -g45
-t15'') \cite{Delcher:2007}. Thirdly, all genomes were translated into
all possible amino acid sequences of length 15 or more and scanned for
homologs of entries in the Pfam database of protein families using
hmmsearch (v3.1dev and the parameter ``--cut\_ga'')
\cite{Punta:2012,Finn:2014}.

\subsection*{Identification of novel RNAs}

From the mapped RNA-seq data, potential novel RNA genes (designated
RNAs of Unknown Function, or RUFs) were picked manually by locating
regions in the genomes that showed high levels of expression without
overlapping annotated protein coding or RNA genes. Only RUFs of
lengths 50 to 400 nucleotides were included, yielding a total of 844
RUFs in Bacteria and 78 RUFs in Archaea.

\subsection*{Homology search and structure prediction}

Homologs of the identified RUFs were found in all the downloaded
genomes using nhmmer \cite{Wheeler:2013} in an iterative fashion:
First, the RUF sequence alone was used in the scan; then, all hits
with E-value~$<0.001$ were included and a HMM built. This was iterated
5 times. The alignments from the RUF homology search were analyzed
further by investigating the potential for secondary structure
formation using RNAz \cite{Gruber:2010} and alifoldz
\cite{Washietl:2004}. Protein coding potential of the RUFs was
assessed using RNAcode \cite{Washietl:2011}. Overlaps between
potential RUF homologs in other strains/species and all the annotations in the
respective genomes were also assessed.

\subsection*{Comparative expression and conservation analysis}

For each strain, the available RNA-seq datasets were pooled and a
list was created of transcripts showing expression in that strain in
at least one experiment (defined as a transcript having a median depth
of at least 10 reads in any experiment). A RUF homolog was defined as
being expressed if the median read depth of the homologous region was
at least 10X. Transcripts were ranked for each strain based on median
expression (i.e. the most highly expressed transcript will have rank
1), which makes relative comparison across strains and datasets
possible. The final set comprises 452 different Rfam families, 922
different RUFs, and 7249 different Pfam domains.

For comparative analysis, if a gene was found to be expressed in more
than one strain/species, the minimum rank was used (i.e. showing the
relatively most abundant expression of the gene). This ensures that
transcripts that are always low abundance will remain low abundance,
whereas genes that are highly abundant in at least one of the sampled
time points and conditions will be treated as such. The ranking is
used as a measure of expression.

``Family conservation'' is based on SSU rRNA alignments of
all Bacteria and Archaea, respectively. For each genome, the best hit
to the Rfam model of SSU rRNA was extracted (RF00177 for Bacteria and
RF01959 for Archaea). The sequences were aligned to the model using
cmalign \cite{Nawrocki:2013}. Finally, a distance matrix was
calculated using dnadist \cite{FELS05} with the F84 model
\cite{Kishino:1989,Felsenstein:1996} which allows for different
transition/transversion rates and for different nucleotide
frequencies. The pairwise strain/species distances produced in this manner
estimate the total branch length between any pair of strains/species. For any
gene found in two or more strains/species, the maximum pairwise distance is
used as the conservation score. Upper and lower quartiles of the
distributions are used to define sets of high, medium and low
expression and conservation, respectively. (Expression, upper
quartile: 204. Expression, lower quartile: 1660. Conservation, upper
quartile: 0.478. Conservation, lower quartile: 0.267).

\subsection*{Quality control of RNA-seq datasets}

We ranked datasets based on the following quality control metrics
(values reported in Table S3).
	
{\bf Strand correlation:} We calculated correlation between the reads
on the two strands. If the dataset is unstranded, we expect a
correlation close to 1.
	
{\bf Expression of core genes:} We defined a set of 40 core
protein-coding genes based on \cite{Wu:2013,Darling:2014} and 16 noncoding RNA
genes (the union of tRNA, RNaseP, tmRNA, SRP, 6S and rRNA RNA
families) \cite{Gardner:2011,Burge:2013}. If the median read depth is
greater than 10X, we defined the gene as expressed. For each dataset,
we report the fraction of the core genes that are expressed.
	
{\bf Coverage:} We calculated coverage as the fraction of the
genome covered by at least 10 mapped reads.
	
{\bf Fraction mapped reads:} For each dataset, we ascertained the
fraction of mapped reads.
	
{\bf Concordance:} To measure how well a given RNA-seq dataset
corresponds to the annotated genes in a genome, we developed a
concordance metric. For this, we define true positives (TP) to be the
number of annotated positions that are expressed; false positives (FP)
to be the number of unannotated positions that are expressed; true
negatives (TN) to be the number of unannotated positions that are not
expressed; and false negatives (FN) to be the number of annotated
positions that are not expressed. Note, not all annotated genes are
expected to be expressed, and not all unannotated positions are
false. Therefore, we calculate the positive predictive value (PPV):
\[
PPV=\frac{TP}{TP+FP}
\]
	
This measures the fraction of expressed positions that are
annotated. We also calculate the fraction of the genome that is
annotated:
\[
ANN=\frac{TP+FN}{TP+FP+TN+FN}
\]

To make the PPV more robust, our final concordance metric normalizes
PPV by ANN.


\section*{Acknowledgments}

SL is supported by a Marie Curie International Outgoing Fellowship
within the 7th European Community Framework Programme. SUU is
supported by a Biomolecular Interaction Centre and Bluefern
Supercomputing Facility joint PhD Scholarship from the University of
Canterbury. LB is supported by a Research Fellowship from the
Alexander von Humboldt Stiftung/Foundation. NRT was supported by the
Wellcome Trust (grant number 098051). AMP \& PPG are both supported by
Rutherford Discovery Fellowships, administered by the Royal Society of
New Zealand.

\bibliography{bibliography}

\newpage
\clearpage

\section*{Supplementary Table Legends}

\noindent
{\bf Table S1:} The Pfam, Rfam and RUF identifiers for each entry
corresponding to Figure~\ref{fig:1}.

\noindent
{\bf Table S2:} Strain/species names, genome accessions, RNA-seq data
sources, Pubmed IDs, sequencing platform and notes for each dataset
used for this study.

\noindent
{\bf Table S3:} Quality control measures computed for each RNA-seq
dataset used in this study. The values are defined in detail in the
Methods section.

\end{document}